\documentclass[letter,12pt]{article}

\usepackage{pstricks}
\usepackage{epsfig}
\usepackage{amsmath}
\usepackage{bm}
\usepackage{array}
\usepackage{subfig}
\usepackage{graphicx}
\usepackage{cite}
\usepackage{amssymb}

\def\({\left(}
\def\){\right)}

\begin{document}
\setlength{\baselineskip}{18pt}
\vspace{-3cm}
\begin{flushright}
OSU-HEP-12-06\\
UMD-PP-012-009
\end{flushright}

\renewcommand{\thefootnote}{\fnsymbol{footnote}}
\vspace*{0.3in}

\begin{center}
{\Large\bf Coupling Unification, GUT--Scale Baryogenesis and \\[0.05in]
Neutron--Antineutron Oscillation in \boldmath{$SO(10)$}}
\end{center}

\vspace{0.3cm}
\begin{center}
{ \bf {}~K.S. Babu}$^a$\footnote{Email:
babu@okstate.edu} and {\bf R.N. Mohapatra}$^b$\footnote{Email: rmohapat@umd.edu}
\vspace{0.3cm}

{\em $^a$Department of Physics, Oklahoma State University,
Stillwater, OK 74078, USA }
\vspace*{0.3cm}

{\em $^b$Maryland Center for Fundamental Physics, Department of Physics,\\ University of Maryland,
College Park, MD 20742, USA }

\end{center}

\setlength{\baselineskip}{18pt}
\thispagestyle{empty}

\begin{abstract}
We show that unification of the three gauge couplings can be realized
consistently in a class of non-supersymmetric $SO(10)$ models with a
one--step breaking to the Standard Model if a color--sextet
scalar field survives down to the TeV scale. Such scalars, which should be
accessible to the LHC for direct detection, arise naturally in $SO(10)$
as remnants of the seesaw mechanism for neutrino masses. The diquark
couplings of these scalars lead to $\Delta B = 2$ baryon number violating
processes such as neutron--antineutron oscillation. We estimate the
free neutron--antineutron transition time to be $\tau_{n-\overline{n}}
\approx (10^9-10^{12})$ sec., which is in the interesting range for next
generation $n-\overline{n}$ oscillation experiments.  These models also realize naturally the
recently proposed $(B-L)$--violating GUT scale baryogenesis which survives to low
temperatures unaffected by the electroweak sphaleron interactions.

 \end{abstract}

\newpage
\renewcommand{\thefootnote}{\arabic{footnote}}
\setcounter{footnote}{0}

\section{Introduction}

Understanding the nature of TeV scale physics beyond the Standard Model (SM) has been a
major focus of research in particle physics for many years. It has
acquired a new sense of urgency in view of the discovery potential of the CERN
Large Hadron Collider (LHC) experiments.  Ongoing non-accelerator and low energy
experiments searching for
dark matter, neutrinoless double beta decay, rare muon decays, and the supersymmetric modes of
proton decay are also sensitive to TeV scale physics, although only indirectly.
On the theoretical front the idea of grand unification, which
provides a framework for unifying the various forces of Nature and  the
different types of matter (quarks and leptons), remains to be one of the
leading candidates for physics beyond the SM.  Grand unified theories (GUTs) based on $SO(10)$ gauge
symmetry provide arguably the most natural framework for small neutrino masses.
A class of minimal $SO(10)$ models which employs a {\bf 126}-plet Higgs field \cite{babu} to break the
baryon number minus lepton number $(B-L)$ gauge symmetry
has been found to be quite successful \cite{neutrino} in explaining the
observed neutrino mixing pattern, including the
recent measurements of $\sin^2 2\theta_{13}$ \cite{t2k,minos,doublechooz,dayabay,reno}.

Despite their elegance and success in explaining the pattern of neutrino mixing and fermion masses in general, the only
sure way to test GUTs  appears to be the discovery of proton decay into specific final states.
The most widely studied decay modes are $p\to e^+\pi^0$ and
$p\to \bar{\nu} K^+$, with the neutrino mode arising prominently in supersymmetric GUTs.
These modes have the feature that they preserve  $(B-L)$ symmetry \cite{weinberg1}.
While this is a generic feature of most minimal single step GUT models, it was recently pointed
out\cite{babu1} that since $SO(10)$ models break $(B-L)$ gauge symmetry (necessary for generating
Majorana neutrino masses), they admit scenarios where
$(B-L)$--violating nucleon decay modes \cite{weinberg2} such as $n\to e^-\pi^+$
arise at an observable level.  These modes, which obey the selection rule $\Delta(B-L) = -2$,
become significant if we choose alternative coupling unification routes.
We continue to explore this theme in the present paper, focusing on $\Delta(B-L) = -2$ processes more generally.
We first point out that in the presence of a color sextet scalar field transforming as $(6,1,1/3)$ under
$SU(3)_C \times SU(2)_L \times U(1)_Y$ (denoted as $\Delta_{u^cd^c}$), along with a complex weak triplet
$(1,3,0)$ at the TeV scale, the three gauge couplings of the Standard Model unify at a scale $M_U \approx
10^{15.5}$ GeV.  If the $\Delta_{u^cd^c}$ field is replaced by the color sextet scalar $\Delta_{d^cd^c}$ transforming
as $(6,1,-2/3)$, the success in the unification of couplings is not altered by much.  These color sextet scalars
arise naturally in $SO(10)$ as the partners of the Higgs field employed for the seesaw mechanism.  Specifically,
$\Delta_{u^cd^c}$ and $\Delta_{d^cd^c}$ are contained in the ${\bf 126}$--plet of Higgs in $SO(10)$.  Our results
provide added motivation for the search for these color sextet scalars at the LHC.

The presence of either $\Delta_{u^cd^c}$ or $\Delta_{d^cd^c}$ color sextet field at the TeV scale will induce
neutron--antineutron oscillation, which is a  $\Delta(B-L) = -2$ process.  It turns out that this
process can probe $(B-L)$ violation occurring at energies all the way up to the unification scale of order $10^{15}$ GeV, given that
one of the color sextet scalars has a TeV scale mass.
We estimate the transition
time for free neutron oscillation to be $\tau_{n-\overline{n}} \approx (10^9-10^{12})$ sec., which is in the observable range of proposed experiments
with currently available reactor neutron fluxes.

In Ref. \cite{babu1} we proposed a new way of generating baryon asymmetry of the universe at the GUT scale in $SO(10)$,
in the $(B-L)$ violating decays of GUT scale scalars.  Such an asymmetry will survive down to low temperatures, unaffected
by the electroweak sphaleron processes, owing to the violation of $(B-L)$ in these decays.  Here we show that this baryogenesis mechanism
is realized naturally in the present context with a TeV scale color sextet.  Suppose that $\Delta_{u^cd^c}$ has a TeV
scale mass, while $\Delta_{d^cd^c}$ has a GUT scale mass. The decay $\Delta_{d^cd^c} \rightarrow \Delta_{u^cd^c}^* \Delta_{u^cd^c}^*$
is then allowed in $SO(10)$, with the relevant vertex arising for example from the $({\bf 126})^4$ quartic coupling, after insertion
of a $(B-L)$ breaking vacuum expectation value (VEV).  Both $\Delta_{u^cd^c}$ and $\Delta_{d^cd^c}$ have $(B-L) = +2/3$, so this
decay violates $(B-L)$ by two units.  We show that there is sufficient CP violation in this decay in minimal $SO(10)$ models,
so that adequate baryon asymmetry
is induced.  Thus in this class of models there is a direct connection between baryon asymmetry of the universe and neutron--antineutron
oscillations.

In Sec. 5 we make some general remarks, not specific to $SO(10)$ models, on light scalars accessible to the LHC and baryon number violation.
We identify scalars that have Yukawa couplings to the fermions which can survive down to the TeV scale without causing problems with rapid nucleon decay.
Allowing for Planck scale suppressed operators, we find that only color singlets, sextets, and octets are safe from rapid nucleon decay
problem.  Color triplet scalars such as $\omega(3,1,-1/3)$ and $\rho(3,2,1/6)$ would have dimension five Planck scale suppressed couplings
to fermions that violate baryon number \cite{barr}.  Such particles are disfavored at the TeV scale, unless there is a new symmetry such
as $(B-L)$ that is broken at or below a scale  $v_{BL} \sim 10^{14}$ GeV, so that the Planck suppressed operators have an
additional $(v_{BL}/M_{\rm Pl})$ suppression factor.  We argue that
models with TeV scale color sextets developed here for successful unification are fully consistent with these general nucleon decay constraints.

\section{Coupling unification with TeV--scale color sextet scalar}

In this section we demonstrate in a bottom--up approach how the presence of a color sextet scalar field
along with a complex $SU(2)_L$ triplet scalar at the TeV scale fixes the problem of unification mismatch of the three gauge couplings
that occurs in the Standard Model. We then show that these fields arise naturally in minimal $SO(10)$ unified theories.  Our unification
analysis uses one--loop renormalization group equations (RGE), and ignores high scale threshold effects.\footnote{The two--loop RGE effects
are expected to be small in non--SUSY theories.  The high scale threshold effects can be significant, causing some uncertainty
in the predictions for proton decay and $n-\overline{n}$ oscillation.} In several variants we find consistent gauge
coupling unification without the need for an intermediate scale.  In non--SUSY $SO(10)$ models the traditional route is to assume
an intermediate symmetry such as $SU(4)_C \times SU(2)_L \times SU(2)_R$ realized above a scale of order $(10^{9} - 10^{14})$ GeV.
Gauge coupling unification in this traditional route has been studied in great detail in Ref. \cite{mohap}.  Here we divert from that route
and assume that the gauge symmetry is simply $SU(3)_C \times SU(2)_L \times U(1)_Y$ all the way to the unification scale.  Thus, in
our framework, $SO(10)$ breaks directly down to the SM, but in the process leaves a color sextet and weak triplet scalars light to the TeV scale.

\begin{figure}[h]
\centering
	\includegraphics[scale=0.7]{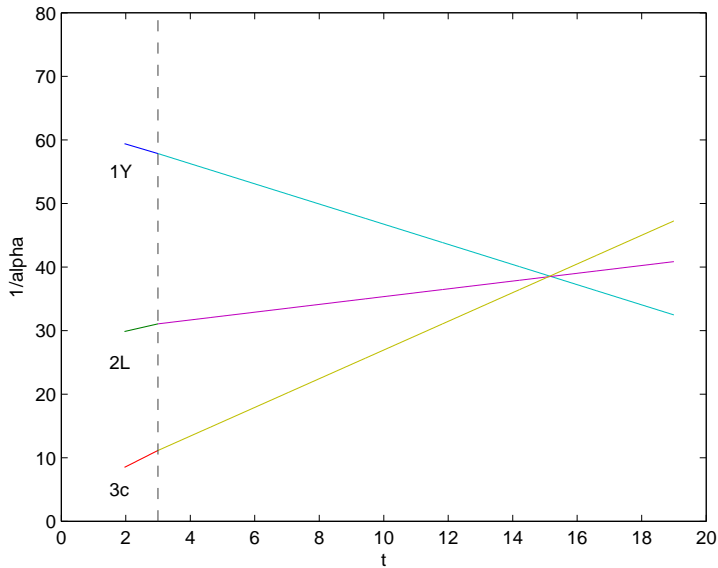}
	\caption{Evolution of the three SM gauge couplings as a function of $t={\rm ln}\mu$ with a color sextet scalar $\Delta_{u^cd^c}(6,1,1/3)$, a complex
 weak triplet scalar $\Delta(1,3,0)$ and a second Higgs doublet $H(1,2,1/2)$ included at 1 TeV.  Results are obtained with one--loop RGE.}
	\label{unif}
\end{figure}

Consider now a scenario where a color sextet scalar field $\Delta_{u^cd^c}$ with the quantum numbers $(6,1,1/3)$ and a complex weak triplet
scalar $\Delta(1,3,0)$ survive down to the TeV scale.  Writing the gauge beta functions as $\beta_i (\alpha_i)= -\frac{b_i}{2\pi}$ ($i$ stands for 1,2,3 depending on whether the gauge group is $U(1)_Y$, $SU(2)_L$ or $SU(3)_C$) so that
the evolution equations for the gauge couplings are given by: $\frac{d \alpha^{-1}_i}{dt}= -\frac{b_i}{2\pi}$, we
find for the scenario mentioned above $(b_1,\, b_2,\, b_3)=(127/30,\, -11/6, \,-37/6)$. If the TeV scale scalar spectrum
also contains a second Higgs doublet $H(1,2,1/2)$, then we have $(b_1,\, b_2,\, b_3)=(13/3,\, -5/3, \,-37/6)$.  We plot
the evolution of the three inverse gauge couplings $\alpha_i^{-1}$ in Fig. \ref{unif} for this scenario.  Excellent unification is
found to occur without any intermediate scale. Here we kept the masses of all scalars to be at 1 TeV.  We note that even with
a single Higgs doublet in the low energy theory, very good unification is found to occur.

We have also studied the case where a $\Delta_{d^c d^c}$ color sextet scalar with the quantum number $(6,1,-2/3)$ survives to the TeV sale,
along with a complex $\Delta(1,3,0)$ scalar and a second Higgs doublet $H(1,2,1/2)$.  This scenario would correspond to the beta function coefficients being $(b_1,\, b_2,\, b_3)=(71/15,\, -5/3, \,-37/6)$, which differs only very slightly from the $\beta$ functions that go into Fig. \ref{unif}.
Excellent results are obtained for coupling unification in this case as well\footnote{The fact that color sextets help non-susy coupling unification was also noted in~\cite{patel}. We thank K. Patel for bringing this work to our attention.}.

It is interesting to note that a unification triangle would begin to develop as the masses of the TeV scale scalars are increased
beyond $2-3$ TeV. While the area of this triangle is much smaller than the corresponding one that arises purely within the SM, and unification
acceptable once high scale threshold effects are included, it is nevertheless encouraging that the preferred mass of the color sextet is around
1 TeV, which is then accessible to direct detection at the LHC \cite{LHC}.

From Fig. \ref{unif} we infer the unification scale to be near $3 \times 10^{15}$ GeV, and the unified gauge coupling to be $\alpha_X \approx
1/40$.  The lifetime for gauge boson mediated proton decay $p \rightarrow e^+ \pi^0$ is close to the present experimental limit of $\tau (p \rightarrow
e^+ \pi^0) > 1.2 \times 10^{34}$ yrs. \cite{superk} in this model as we show below.  The rate for proton decay into this mode is given by
\begin{equation}
\Gamma(p \rightarrow e^+ \pi^0) \approx \frac{m_p}{16\pi f_\pi^2}(1+D+F)^2 \alpha_H^2\left(\frac{g_X^2 A_R}{M_X^2}\right)^2~.
\end{equation}
Here $D\simeq 0.8$ and $F\simeq 0.47$ are chiral Lagrangian factors, $\alpha_H \simeq 0.012~{\rm GeV}^3$ is the hadronic matrix element,
$A_R \simeq 3.8$ is the renormalization factor of the effective dimension six operator,
and $M_X$ is the mass of the $X$ and $Y$ gauge bosons of $SO(10)$.  (Here we assume that the $X'$ and $Y'$ gauge bosons of $SO(10)$
which lie outside of the $SU(5)$ subgroup and mediate this decay are somewhat heavier than the $(X,Y)$ gauge bosons.)  With the values
of these parameters as quoted, we find $\Gamma^{-1}(p \rightarrow e^+ \pi^0) \approx 2.3 \times 10^{34}$ yrs. if $M_X = 6 \times
10^{15}$ GeV is assumed.  Allowing for high scale threshold effects which may raise $M_X$ by a factor of 2 or so, we infer that
the lifetime for this mode is typically less than about $10^{35}$ yrs., which is currently being probed by the SuperKamiokande experiment.

How do these light scalars fit into the framework of non--SUSY $SO(10)$?  We note that minimal models assume a ${\bf 126}$--plet of
Higgs for breaking $(B-L)$ symmetry and for generating neutrino masses.  The ${\bf 126}$ decomposes under $SU(2)_L \times SU(2)_R
\times SU(4)_C$ subgroup of $SO(10)$ as
\begin{equation}
126 = (1,1,6) + (3,1,10) + (1,3, \overline{10}) +(2,2,15)~.
\end{equation}
It is the $(1,3,10)$ fragment of the $\overline{\bf 126}$ that generates a large Majorana mass for the right--handed neutrino,
after acquiring a GUT scale VEV along its SM singlet component, thus breaking $(B-L)$ by two units.  The same $(1,3,10)$ fragment
contains the color sextet fields $\Delta_{u^cd^c}(6,1,1/3)$, $\Delta_{d^cd^c}(6,1,-2/3)$ and $\Delta_{u^cu^c}(6,1,4/3)$, with
the quantum numbers referring to $SU(3)_C \times SU(2)_L \times U(1)_Y$.  Thus the $\Delta_{u^cd^c}$ color sextet, or the
$\Delta_{d^cd^c}$ color sextet that we assumed to survive down to the TeV scale arises naturally in the process of implementing
the seesaw mechanism in $SO(10)$.  In order to complete the symmetry breaking an additional ${\bf 54}$ and/or ${\bf 45}$ Higgs field
is needed.  These fields decompose under  $SU(2)_L \times SU(2)_R
\times SU(4)_C$ as
\begin{eqnarray}
54 &=& (1,1,1) + (3,3,1) + (1,1,20') + (2,2,6) \nonumber \\
45 &=& (3,1,1) + (1,3,1) + (1,1,15) + (2,2,6)~.
\label{decom}
\end{eqnarray}
The $(3,3,1)$ fragment of ${\bf 54}$ and the $(3,1,1)$ fragment of ${\bf 45}$ contain the $\Delta(1,3,0)$ field that we assumed to
survive to the TeV scale.  We realize that making these fields light would entail additional fine--tunings.

\section{Neutron--antineutron oscillation in \boldmath{$SO(10)$} }

Neutron-antineutron oscillation arises in theories beyond the SM from $B = \pm 2$ dimension nine effective operators \cite{mohapreview}.
There are several such operators, we focus on the operator  $u^cd^cd^cu^cd^cd^c$, which is the one induced by the color sextet fields used to fix the unification problem of the Standard Model in the previous section.  Due to its high dimensionality, this and other such
operator scale as $M^{-5}$, where $M$ is a heavy mass scale
corresponding to the scale of baryon number violation.
Very roughly we can estimate the $n-\overline{n}$ oscillation time to be $\tau_{n-\overline{n}} \approx \frac{ M^5}{\Lambda^6_{QCD}}$.
The present limit on neutron--antineutron oscillation inferred from constraints on matter disintegration is  $\tau_{n-\bar{n}}\geq 2\times 10^8$ sec.
\cite{SuperK1}.  The direct limit from free neutron oscillation search is $\tau_{n-\overline{n}} > 10^8$ sec. \cite{ill}, which is slightly
weaker, but does not have the nuclear uncertainties that affect the limit derived from matter disintegration.  These limits translate to a lower bound
on the heavy mass scale $M\geq 10^{5.5}$ GeV. Generally, in seesaw extensions of the Standard Model with local $(B-L)$ symmetry, we expect that $M\sim v_{BL}$\cite{marshak}, leading to the general perception that in unified $SO(10)$ models for neutrino masses, the GUT scale breaking of
$(B-L)$ symmetry will yield infinitesimally tiny and hence unobservable strength of $n-\bar{n}$ oscillation amplitude. Although this naive estimate
is off by a factor of 100 when compared to more detailed model calculations including the hadronic matrix element \cite{sarkar}, the
main conclusion does not get altered.

The outcome of the operator analysis changes drastically however, once there are new particles at the TeV scale.
For example, in a theory with a color sextet Higgs field $\Delta_{u^cd^c}$ at the TeV or sub--TeV scale as in the previous
section, one can write down a new $(B-L)=\pm 2$  effective operator $\Delta^*_{u^cd^c}\Delta^*_{u^cd^c}
d^cd^c + h.c.$ (where we have defined the $B-L$ of the scalar field $\Delta_{u^cd^c}$ to be $+2/3$ via its Yukawa coupling $u^cd^c\Delta_{u^cd^c}$). Note that this operator has only dimension five and therefore scales by one power of the heavy scale $M$.
All other new mass scales appearing in the $d=9$ operator that mediates $n-\overline{n}$ oscillation will be near a TeV.
To see how $n-\bar{n}$ oscillation arises in such theories, recall that the $\Delta_{u^cd^c}$ field belongs to ${\bf 126}$--plet Higgs of
$SO(10)$ GUT.  There are other fields in the ${\bf 126}$ with GUT scale masses, e.g., $\Delta_{d^cd^c}$ which has the Yukawa
coupling $d^cd^c \Delta_{d^cd^c}$.  Of course, these
Yukawa couplings by themselves would  preserve baryon number.  The Higgs potential contains a quartic coupling $\lambda ({\bf 126})^4$, which when expanded generates a baryon number violating
cubic scalar couplings $\lambda v_{BL} \, \Delta_{d^cd^c}\Delta_{u^cd^c}\Delta_{u^cd^c}$, where $v_{BL}$ is the $B-L$ breaking vacuum
expectation value of the SM singlet component in ${\bf 126}$. A combination of these interactions violates baryon number and generates
$n-\overline{n}$ oscillation.

\begin{figure}[h]
\centering
	\includegraphics[scale=0.6]{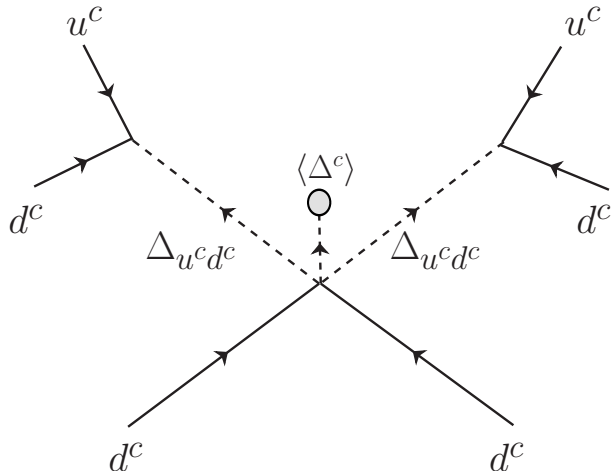}
	\caption{$n-\overline{n}$ oscillation amplitude from the effective interaction of Eq. (\ref{nnbar}).}
	\label{nnbar}
\end{figure}

At the $SO(10)$ level, the relevant interactions for $n-\overline{n}$ oscillation is given by
\begin{eqnarray}
{\cal L}_{\Delta B \neq 0} &=& f_{dd} \,d^c d^c \Delta_{d^cd^c} + \frac{f_{ud}}{\sqrt{2}} \,(u^c d^c + d^c u^c) \Delta_{u^cd^c} + f_{uu}\, u^c u^c \Delta_{u^cu^c} \nonumber \\
&+& \lambda v_{BL}\, \left(\Delta_{u^c d^c} \Delta_{u^c d^c} \Delta_{d^c d^c} + \Delta_{d^c d^c} \Delta_{d^c d^c}
\Delta_{u^cu^c} \right) + h.c.
\end{eqnarray}
In the symmetric limit of $SO(10)$ we have $f_{uu} = f_{dd} = f_{ud}$ with these matrices being symmetric in family indices (which are
not shown explicitly).  Now suppose that $\Delta_{u^cd^c}$ has a TeV scale mass, while $\Delta_{d^cd^c}$ and $\Delta_{u^c u^c}$ have
GUT scale masses.  The effective $B$--violating interactions below the GUT scale is then given by
\begin{equation}
{\cal L}_{\Delta B \neq 0}^{\rm eff} = \frac{f_{ud}}{\sqrt{2}} \,(u^c d^c + d^c u^c) \Delta_{u^cd^c} + \frac{\lambda v_{BL}f_{dd}}{M_{\Delta_{d^cd^c}}^2} \,d^c d^c \, \Delta_{u^cd^c}^* \Delta_{u^cd^c}^* + h.c.
\label{eff}
\end{equation}
Note that the second term in Eq. (\ref{eff}) has one power of an inverse mass, since $\lambda v_{BL}$ and $M_{d^cd^c}$ are
both of the same order.  This effective interaction induces through Fig. \ref{nnbar} an amplitude for $n-\overline{n}$ oscillation.

The amplitude for  $n-\overline{n}$ oscillation arising from Fig. \ref{nnbar} is given by
\begin{eqnarray}
G_{N-\bar{N}}~\simeq \frac{\eta f_{ud}^2f_{dd}\lambda v_{BL}}{M^4_{\Delta_{u^cd^c}}M^2_{\Delta_{d^cd^c}}}~.
\end{eqnarray}
Here $\eta$ is a renormalization factor that accounts for the running of the Yukawa coupling $f_{dd}$ and
the effective $d=5$ operator in Eq. (\ref{eff}).  Since $\Delta_{u^cd^c}$ is a color sextet, the RGE
factor for $f_{dd}$ is expected to be relatively large, we estimate it to be of order 4 in going from
the GUT scale to the weak scale.  Similarly, the RGE factor for the effective $d=5$ operator should be of
order 6.  Thus the net RGE factor can be as large as $\eta \approx 50-100$ for the $n-\overline{n}$ effective operator.
(A detailed calculation of the RGE factor will be quite useful.)
Choosing $f_{dd} = f_{ud} = 10^{-3}$, $\eta = 100$, $\lambda v_{BL} = 6 \times 10^{15}$ GeV, $M_{\Delta_{u^cd^c}}
= 500$ GeV, and $M_{\Delta{d^cd^c}} = 10^{14}$ GeV as an example, we find $\tau_{n-\overline{n}} \approx
5 \times 10^{11}$ sec.  For the hadronic matrix element that takes six quarks to neutrons,
we have used a value $10^{-4}$ GeV$^6$ \cite{sarkar}.
Here the choice of $f_{dd}$ and $f_{ud}$ corresponds to the first family coupling.
 If these couplings are $5 \times 10^{-3}$ instead, we would obtain $\tau_{n-\overline{n}} \approx 4 \times 10^{9}$ sec.
We note that the mass of the $\Delta_{d^cd^c}$ scalar being somewhat below the unification scale goes well and
may even be required in the model in order for generating the right amount of baryon asymmetry, which is discussed
in the next section.  We conclude here that $n-\overline{n}$ oscillation time can be within reach of proposed
experiments with existing luminosity of neutron beams.  With a TeV scale color sextet, $n-\overline{n}$ experiments
can probe baryon number violation all the way to the GUT scale.  While there are significant uncertainties in the
oscillation time, we estimate that $\tau_{n-\overline{n}} = (10^9 - 10^{12})$ sec. to be very natural in the model.

\section{GUT scale baryogenesis from color sextet decays}

In a recent paper \cite{babu1} we pointed out that in $SO(10)$ models, breaking of $B-L$ symmetry to generate neutrino masses via the seesaw mechanism allows for a new scenario for baryognesis.  It was shown that an asymmetry in $B-L$ can be generated via the decay of
color triplet scalars.  This asymmetry is not erased by the electroweak sphaleron interactions since sphalerons conserve $B-L$ symmetry.
This mechanism can be implemented in the context of color sextet scalars developed in this paper, as we show below.

We consider the decay of the color sextet scalar $\Delta_{d^cd^c}$ which is assumed to have a GUT scale mass.
We also assume, as in the previous sections, that the color sextet scalar $\Delta_{u^cd^c}$ has a TeV sale mass.
$\Delta_{d^cd^c}$ has two classes of decay modes:  $\Delta_{d^cd^c}\to \overline{d^c}\,\overline{d^c}$ and $\Delta_{d^cd^c} \rightarrow  \Delta^*_{u^cd^c}\Delta^*_{u^cd^c}$. The first decay mode, along with the decay of $\Delta_{u^c d^c} \to \overline{u^c}\,\overline{d^c}$,
assign $(B-L) = +2/3$ for the two color sextets.  The decay $\Delta_{d^c d^c} \to \Delta_{u^c d^c}^* \Delta_{u^c d^c}^*$ would then violate
$B-L$ by two units.  If there is an asymmetry in the decay rates $\Delta_{d^cd^c} \to \Delta_{u^c d^c}^* \Delta_{u^c d^c}^*$ and the
CP conjugate process $\Delta_{d^cd^c}^* \to \Delta_{u^c d^c} \Delta_{u^c d^c}$, a net baryon asymmetry can be generated.  This CP
asymmetry arises through the interference between tree--level and one--loop diagrams.

We now show that in minimal $SO(10)$ models there is a source for CP asymmetry in the $B-L$ violating decays of $\Delta_{d^c d^c}$.
As already noted, to complete symmetry breaking of $SO(10)$ a Higgs field such as a ${\bf 54}$ is needed apart from the ${\bf 126}$.
Adopting a single ${\bf 126}$ and a single ${\bf 54}$ Higgs fields for $SO(10)$ breaking, we find that there are two color sextet
fields of the type $\Delta_{d^cd^c}(6,1,-2/3)$.  One of them is from the ${\bf 126}$ as already discussed (denoted as $\Delta_{d^cd^c}(126)$), while the other one arises
from the ${\bf 54}$.  The $(1,1,20')$ fragment of ${\bf 54}$ (see Eq. (\ref{decom})) contains this field, which we denote as $\Delta_{d^c d^c}(54)$.
Now there are terms in the Higgs potential which mix these two fields, for example the cubic coupling $\mu\, ({\bf \overline{126}})^2\, {\bf 54} + h.c.$ With such mixings, the mass eigenstates become
\begin{eqnarray}
\Delta_{d^c d^c} &=& a \, \Delta_{d^c d^c}(126) + b \,\Delta_{d^c d^c}(54) \nonumber \\
\Delta'_{d^c d^c} &=& -b^*\, \Delta_{d^c d^c}(126) + a^* \,\Delta_{d^c d^c}(54)
\end{eqnarray}
with $|a|^2 + |b|^2 = 1$, with $a,b$ being combinations of Higgs potential parameters.  The trilinear coupling $\mu\, ({\bf \overline{126}})^2\, {\bf 54}$ and the quartic coupling $\lambda ({\bf 126})^4$
will lead, upon inserting the $B-L$ breaking VEV of the SM singlet in ${\bf 126}$, effective baryon number violating trilinear couplings
given by\footnote{Such trilinear couplings can also receive contributions from quartic couplings of the type $({\bf 126})^2 ({\bf 54})^2$,
which we ignore here for simplicity in presentation.}
\begin{equation}
V^{(3)} = \Delta_{u^c d^c} \Delta_{u^c d^c}\left\{ \Delta_{d^c d^c} (\lambda v_{BL} a^* + \mu b^*) + \Delta'_{d^cd^c}(-\lambda v_{BL} b + \mu a)
\right\} + h.c.
\end{equation}
The fermion Yukawa couplings of $\Delta_{d^c d^c}(126)$ become, in the mass eigenstates of the color sextet scalars,
\begin{equation}
{\cal L}_{\rm Yuk} = f_{dd}\,d^c d^c \,\left(a^* \,\Delta_{d^cd^c} - b\,\Delta'_{d^c d^c}\right) + h.c.
\end{equation}

\begin{figure}[h]
\centering
	\includegraphics[scale=0.5]{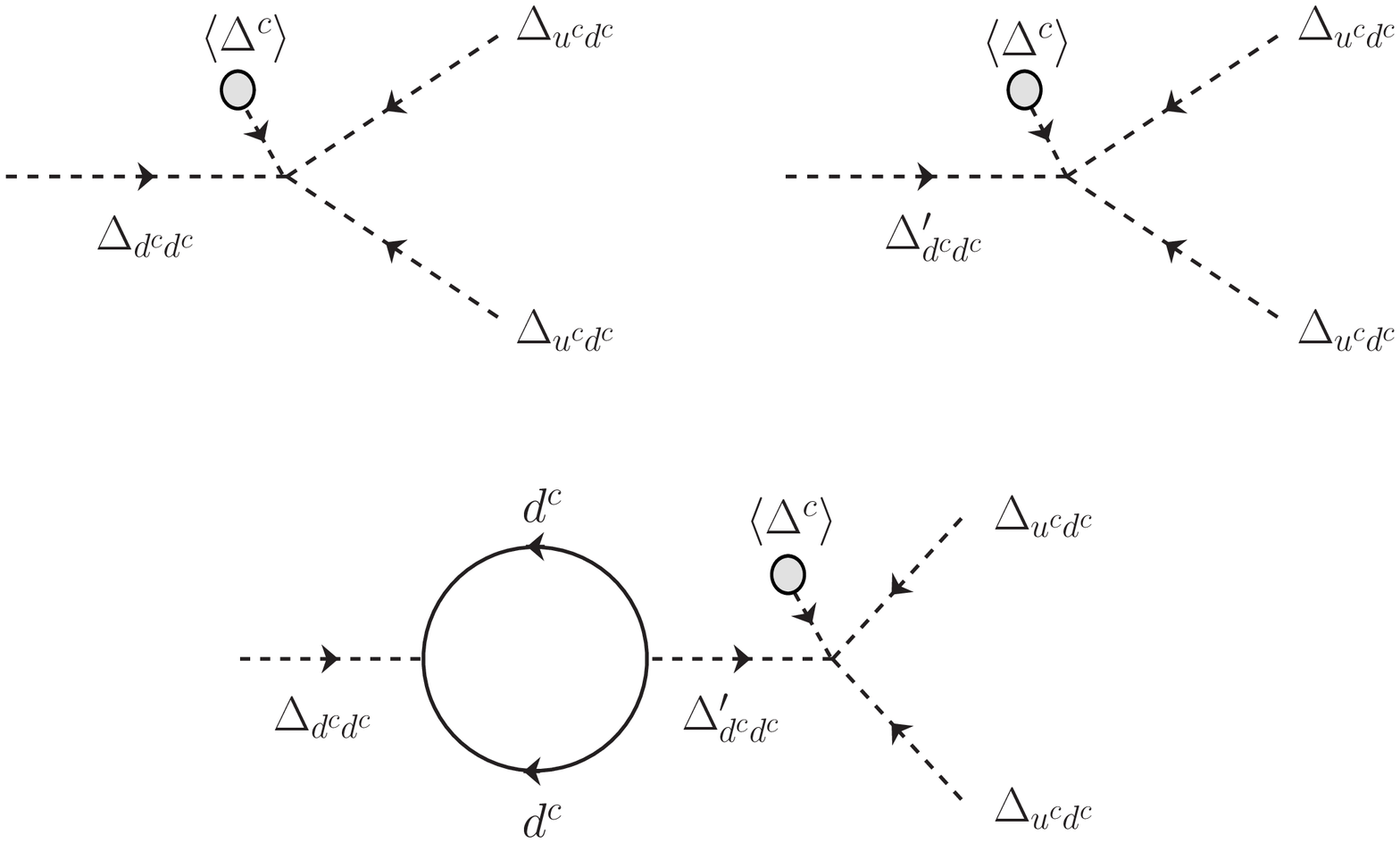}
	\caption{Tree--level diagram and one--loop correction for the decay of the color sextet $\Delta_{d^cd^c}$.}
	\label{decay}
\end{figure}

Consider now the decay $\Delta_{d^cd^c} \to \Delta_{u^c d^c}^* \Delta_{u^cu^c}^*$.  The tree--level diagram and the one--loop
correction to this decay are shown in Fig. \ref{decay}.  The loop diagram is a wave function correction involving the transition of $\Delta_{d^cd^c}$
into the heavier $\Delta'_{d^cd^c}$.  The $B-L$ asymmetry generated in the decay of a $\Delta_{d^cd^c}$ and its conjugate
$\Delta_{d^cd^c}^*$ can be computed as follows.  Let $r$ be the branching ratio for the decay $\Delta_{d^cd^c} \to \Delta_{u^c d^c}^* \Delta_{u^cu^c}^*$ and let $\overline{r}$ be the branching ratio for the conjugate decay
$\Delta_{d^cd^c}^* \to \Delta_{u^c d^c} \Delta_{u^cu^c}$.  The $B-L$ asymmetry generated in the decay of a pair of $\Delta_{d^c d^c}$
and $\Delta_{d^c d^c}^*$  is then given by
\begin{equation}
\epsilon_{B-L} = (r-\overline{r})(B_1-B_2)
\end{equation}
where $B_1 = -4/3$ and $B_2 = 4/3$ are the final state $B-L$ quantum numbers in these decays.  From Fig. \ref{decay}
we find $\epsilon_{B-L}$ to be
\begin{equation}
\epsilon_{B-L} = \frac{2}{\pi} \,{\rm Tr}(f_{dd}^\dagger f_{dd})\, {\rm Im} \left\{
\frac{-\lambda v_{BL} b + \mu a }{\lambda v_{BL} a^* + \mu b^*}  \right\} \left(\frac{x}{1-x}  \right) {\rm Br}
\end{equation}
where
\begin{equation}
x= M_{\Delta_{d^cd^c}}^2/M_{\Delta'_{d^cd^c}}^2
\end{equation}
and where Br stands for the branching ratio for the decay $\Delta_{d^cd^c} \to \Delta_{u^c d^c}^* \Delta_{u^cu^c}^*$.
The baryon number normalized to the entropy is given by
\begin{equation}
\eta = \frac{n_B}{s} = \frac{\epsilon_{B-L}}{g^*} d
\end{equation}
where $g^* \simeq 120$ is the effective number of relativistic degrees of freedom at the time of decay, and $d$ is
the dilution factor.  We see that for $f \simeq 1$ (corresponding to the third family Yukawa coupling), $M_{\Delta_{d^cd^c}} \approx
M_{\Delta'_{d^cd^c}}/10$, $\eta_B \approx 10^{-10}$ can be generated, along the same line as in Ref. \cite{babu1}.\footnote{For alternative
scenarios of high scale baryogenesis see \cite{maekawa}.}

\section{Constraints on TeV scale colored particles from nucleon decay}

In this section we make some general remarks, not specific to $SO(10)$ GUTs, on scalars that have non-negligible Yukawa couplings to the SM fermions and which survive
down to the TeV scale.  Constraints from baryon number violation would disfavor several such scalars, provided that the Standard Model gauge symmetry
extends all the way to the Planck scale with no new structure in between.  To see this, note that
any  such scalar should transform as a color singlet, color octet, color triplet (or anti-triplet),
or a color sextet (or anti-sextet).  The color singlet and color octet scalars do not violate baryon number and are safe from
nucleon decay constraints.  This is not so for the color triplet scalars, even if one assumes global baryon number symmetry to prevent
rapid proton decay via $d=6$ operators.  Planck scale suppressed $d=7$ operators will be induced, which would typically lead to
rapid nucleon decay.  We illustrate this with explicit examples.

\begin{figure}[h]
\centering
	\includegraphics[scale=0.5]{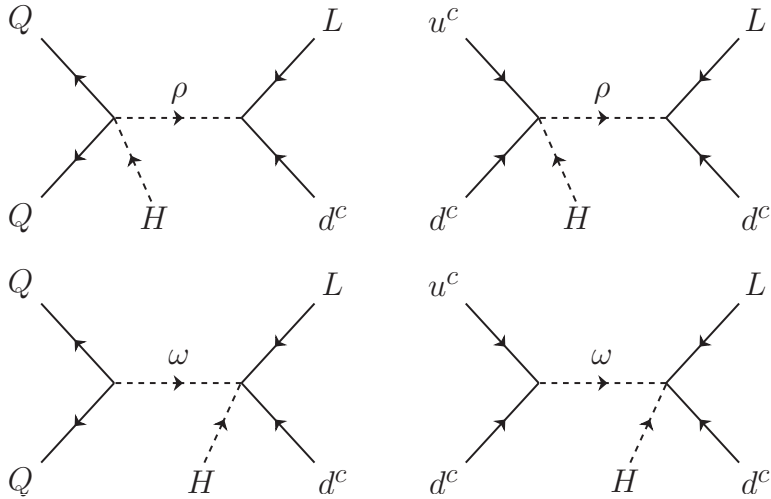}
	\caption{$d=7$ nucleon decay diagrams induced by light color triplet states.}
	\label{rapid}
\end{figure}

There are six types of color triplet (anti-triplet) scalars that can have Yukawa couplings to the fermions \cite{volkas,strumia,dorsner,barr,babu1}.
These have $SU(3)_C \times SU(2)_L \times U(1)_Y$ quantum numbers given by $\rho(3,2,1/6)$, $\omega(3,1,-1/3)$, $\eta(3,1,2/3)$, $\Phi(3,3,-1/3)$,
$\chi(3,2,7/6)$ and $\delta(3,1,-4/3)$ and their conjugates (in the notation of Ref. \cite{babu1}).  Now suppose that the color triplet $\rho(3,2,1/6)$ survives
down to the TeV scale.  LHC can then directly detect this state.  $\rho$ has the Yukawa coupling
\begin{equation}
{\cal L}_\rho = f_\rho L d^c \rho + h.c.
\label{rho1}
\end{equation}
This coupling of course does not violate baryon number (or lepton number), as we can assign $B=1/3, L = -1$ to $\rho$.  Now it turns out
that all symmetries allow the existence of the $d=5$ coupling, suppressed by a high mass scale $M_*$,
\begin{equation}
{\cal L}_\rho^{d=5} = \frac{Q Q \rho H^* }{M_*} + \frac{u^c d^c \rho^* H}{M_*}
\label{rho2}
\end{equation}
where $H(1,2,1/2)$ is the SM Higgs doublet.  Combining the interactions of Eqs. (\ref{rho1})-(\ref{rho2}), we obtain the $d=7$
nucleon decay operator \cite{weinberg1,babu1,barr} as shown in Fig. \ref{rapid}.  The effective $d=7$ operators
$(QQ)^* (L d^c) H$ and $(u^c d^c)(L d^c)H$ have coefficients $f_\rho/(M_* M_\rho^2)$.  The resulting rate for nucleon decay
is estimated to be
\begin{equation}
\Gamma(n \to e^- \pi+) \approx \frac{|f_\rho|^2}{64 \pi} (1+D+F)^2 \frac{\beta_H^2\, m_p}{f_\pi^2}\left(\frac{v}{M_* M_\rho^2} \right)^2
\end{equation}
where $\beta_H \simeq 0.012$ GeV$^3$ is the hadronic matrix element, and $v=174$ GeV is the electroweak VEV.  For $f=10^{-3}$,
$M_\rho = 1$ TeV, and $M_* = 1.2 \times 10^{19}$ GeV one finds $\tau(n \to e^- \pi^+) \approx 5 \times 10^{23}$ yrs., which
is excluded by experimental limits.  One could obtain consistent lifetime by taking $f_\rho$ to be extremely tiny, $f_\rho = 10^{-8}$
would result in $\tau(n \rightarrow e^- \pi^+) \approx 5 \times 10^{33}$ yrs.  Such a small value appears to us to be highly unnatural.

There are several solutions to this problem.  First, color triplet states such as $\rho(3,2,1/6)$ need not show up at the LHC.
Second, if such states do show up, consistency with nucleon decay can be achieved by assuming an intermediate scale symmetry
such as $B-L$.  In the presence of such symmetry, the amplitude for $d=7$ nucleon decay will have an additional suppression factor
$v_{BL}/M_*$.  For $f_\rho = 10^{-3}$ as before, with this additional suppression factor we find that $\tau(n \to e^- \pi^0) \approx
5 \times 10^{33}$ yrs. if $v_{BL} \approx 10^{14}$ GeV.  Any value of $v_{BL}$ less than $10^{14}$ GeV would cure the rapid proton
decay problem in this setup.

In Fig. \ref{rapid} we also show the effective $d=7$ nucleon decay operators induced by a TeV scale color triplet $\omega(3,1,-1/3)$.
It leads to the following effective Lagrangian.
\begin{eqnarray}
{\cal L}_\omega = f_\omega Q Q \,\omega + f'_\omega u^c d^c \omega^* + \frac{L d^c \omega H}{M_*}~.
\end{eqnarray}
The effective $d=7$ nucleon decay operator $(Q Q)^*(L d^c)H$ has a coefficient $f_\omega/(M_* M_\omega^2)$ which results in the
same problem as in the case of $\rho(3,2,1/6)$.  In the present case one has to assume global baryon or lepton number so that
there is no rapid nucleon decay arising from $d=7$ operators such as $(QQQL)/M_\omega^2$.  Again the problem with $d=7$ nucleon
decay can be solved by assuming gauged $B-L$ symmetry broken at a scale below $10^{14}$ GeV.  Identical problems arise with
TeV scale $\eta(3,1,2/3)$, $\Phi(3,3,-1/3)$, $\chi(3,2,7/6)$ and $\delta(3,1,-4/3)$ scalar fields. Gauging of $B-L$ would
again bring back consistency with nucleon decay.

We point out that the potential problem with rapid $d=7$ nucleon decay does not arise with TeV scale color sextet fields,
the reason being that the higher dimensional $d=5$ effective interactions of the light color sextets would not involve
any lepton fields~\cite{marshak,barr}.  Thus the color sextets would lead only to processes such as $n-\overline{n}$ oscillation, and not
to nucleon decay, even with the inclusion of Planck scale suppressed operators.

Finally we comment on a constraint on cubic scalar couplings such as $\mu\, \Delta_{u^c d^c}\Delta_{u^c d^c} \Delta_{d^c d^c}$.
If both of the fields $\Delta_{u^c d^c}$ and $\Delta_{d^c d^c}$ are at the TeV scale, the coefficient $\mu$ has to be also of order
a few TeV for consistency.  Such trilinear couplings, as we saw, do not lead to rapid proton decay and so both scalars being of order
TeV mass is consistent.  If $\mu$ is much larger than the masses of these particles, a possible consistency problem arises as noted in Ref. \cite{babu2}. The point can be summarized by transplanting the argument of Ref. \cite{babu2} to the color sextets as follows: At the one loop level, one induces a quartic scalar coupling of the type $(\Delta^\dagger_{u^cd^c}\Delta_{u^cd^c})^2$ with a coefficient $-1/(16\pi^2) (\mu/M_{\Delta_{u^cd^c}})^4$ and similar terms with $\Delta_{d^cd^c}$. The point is the negative sign in front, which implies that unless $M_{\Delta_{u^cd^c}}\geq \mu$, this term will overwhelm the perturbative tree level term of this type and lead to unstable color breaking vacuum.
This implies that if $M_{\Delta_{u^cd^c}}$ and $M_{\Delta_{d^cd^c}}$ are both in the TeV range, their trilinear coupling $\mu$ better be in the same range. This has important implications for theories where the $B-L$ violating mass $\mu$ originates from the GUT scale. In such theories, one expects $M\sim v_{BL}\sim M_U$. Therefore, the above instability problem can be avoided if two of the three color sextets (e.g. $\Delta_{u^cu^c}$,
$\Delta_ {d^cd^c}$) are at the GUT scale and only one color sextet ($\Delta_{u^cd^c}$) is at the TeV scale. Examination of all possible one loop induced term leads to the conclusion that in this case, the loop induced term does not overwhelm the tree level term, thereby keeping the potential from going to negative infinity for large values of the fields. In the case of GUTs, this result can also be seen by looking at the contribution of the trilinear term to the running of the masses of the color sextet fields. Note the similarity of this phenomenon to constraints on $A$--terms in MSSM \cite{raby}. From this discussion, we conclude that when the $B-L$ breaking scale is close to GUT scale, for the induced $B-L$ breaking color sextet trilinear to be consistent with vacuum stability, only one of the three sextets can remain in the TeV scale while the others have to be at the GUT scale. As we saw in Sec. 2, this is also the scenario that leads to gauge coupling unification without supersymmetry.

\section{Conclusions}

We have shown that a TeV scale color sextet scalar, along with a weak triplet, can fix the unification problem that exists in the SM.
The color sextet scalar arises naturally in $SO(10)$ GUTs as the partner of the Higgs field that takes part in the seesaw mechanism.
Such light color sextets are accessible to the LHC experiments, and should be searched for.  With a color sextet at the TeV scale,
neutron--antineutron oscillation can occur at a rate that is within reach of proposed experiments.  We estimate the transition time
for free neutron oscillations to be $(10^9-10^{12})$ sec in realistic $SO(10)$ models.  Observation of $n-\overline{n}$ oscillation
would probe GUT scale physics, provided that there is a color sextet scalar at the TeV scale.  The existence of color sextets at a TeV is
shown to be safe from nucleon decay problems, even allowing for Planck scale suppressed operators.  This is not the case with TeV scale color
triplet scalars, unless there are additional suppression factors arising from intermediate symmetries such as gauged $B-L$.
We have also shown that the GUT scale baryogenesis mechanism that violates $B-L$ and survives to low temperatures without being
erased by the sphalerons can be achieved in $SO(10)$ in the decay of color sextet scalars.  Thus in these theories there is a
strong connection between the origin of matter in the universe and neutron--antineutron oscillations.

\section*{Acknowledgement}

The work of KSB is supported in part the US Department of Energy, Grant Numbers DE-FG02-04ER41306 and that of RNM  is supported
in part by the National Science Foundation Grant Number PHY-0968854.

\end{document}